\documentclass[%
 reprint,
 superscriptaddress,
 amsmath,amssymb,
 aps,
 prl
]{revtex4-2}

\usepackage{graphicx} 
\usepackage{subcaption}
\usepackage{siunitx}
\usepackage{graphicx}
\usepackage{dcolumn}
\usepackage{bm}
\usepackage{booktabs}       
\usepackage{cancel}
\usepackage{ulem}
\expandafter\let\csname equation*\endcsname\relax
\expandafter\let\csname endequation*\endcsname\relax
\usepackage{amsmath}
\usepackage{ragged2e}
\usepackage{cleveref}

\begin{document}
\title{Probing Nuclear Excitation by Electron Capture in an Electron Beam Ion Trap with Non-destructive Isomer Detection via Precision Mass Spectrometry}

\author{Bingsheng Tu}
\thanks{These authors contributed equally to this work}
\affiliation{Key Laboratory of Nuclear Physics and Ion-Beam Application (MOE), Institute of Modern Physics, Fudan University, 200433 Shanghai, China}

\author{Nan Xue}
\thanks{These authors contributed equally to this work}
\affiliation{Key Laboratory of Nuclear Physics and Ion-Beam Application (MOE), Institute of Modern Physics, Fudan University, 200433 Shanghai, China}
\affiliation{Research Center for Theoretical Nuclear Physics, NSFC and Fudan University, 200438 Shanghai, China}
\affiliation{School of Nuclear Science and Technology,Lanzhou University, 730000 Lanzhou, China}

\author{Jialin Liu}
\affiliation{Key Laboratory of Nuclear Physics and Ion-Beam Application (MOE), Institute of Modern Physics, Fudan University, 200433 Shanghai, China}

\author{Qi Guo}
\affiliation{Key Laboratory of Nuclear Physics and Ion-Beam Application (MOE), Institute of Modern Physics, Fudan University, 200433 Shanghai, China}

\author{Yuanbin Wu}
\email{yuanbin@nankai.edu.cn}
\affiliation{School of Physics, Nankai University, 300071 Tianjin, China}

\author{Zuoye Liu}
\affiliation{School of Nuclear Science and Technology,Lanzhou University, 730000 Lanzhou, China}

\author{Adriana P\'alffy}
\affiliation{University of W\"urzburg, Institute of Theoretical Physics and Astrophysics, Am Hubland, 97074 W\"urzburg,  Germany}

\author{Yang Yang}
\affiliation{Key Laboratory of Nuclear Physics and Ion-Beam Application (MOE), Institute of Modern Physics, Fudan University, 200433 Shanghai, China}

\author{Ke Yao}
\email{keyao@fudan.edu.cn}
\affiliation{Key Laboratory of Nuclear Physics and Ion-Beam Application (MOE), Institute of Modern Physics, Fudan University, 200433 Shanghai, China}

\author{Baoren Wei}
\affiliation{Key Laboratory of Nuclear Physics and Ion-Beam Application (MOE), Institute of Modern Physics, Fudan University, 200433 Shanghai, China}

\author{Yaming Zou}
\affiliation{Key Laboratory of Nuclear Physics and Ion-Beam Application (MOE), Institute of Modern Physics, Fudan University, 200433 Shanghai, China}

\author{Xiangjin Kong} 
\email{kongxiangjin@fudan.edu.cn}
\affiliation{Key Laboratory of Nuclear Physics and Ion-Beam Application (MOE), Institute of Modern Physics, Fudan University, 200433 Shanghai, China}
\affiliation{Research Center for Theoretical Nuclear Physics, NSFC and Fudan University, 200438 Shanghai, China}

\author{Yu-Gang Ma}
\email{mayugang@fudan.edu.cn}
\affiliation{Key Laboratory of Nuclear Physics and Ion-Beam Application (MOE), Institute of Modern Physics, Fudan University, 200433 Shanghai, China}
\affiliation{Research Center for Theoretical Nuclear Physics, NSFC and Fudan University, 200438 Shanghai, China}
\date{\today}

\begin{abstract}
Nuclear excitation by electron capture (NEEC) is an important nuclear excitation mechanism which still lacks conclusive experimental verification. This is primarily attributed to strong background x-/$\gamma$-ray noise and competing nuclear excitation processes which would overshadow the signals in various environments that NEEC takes place. Here, we propose an experimental approach to observe the NEEC process within a background-free environment. Through collisions with a highly-compressed mono-energetic electron beam in an electron beam ion trap, nuclei may get excited to a long-lived isomeric state via the NEEC process. Subsequently, ions can be extracted and Penning-trap mass spectrometry employed to unambiguously detect the isomer. Our study focuses on the promising candidate $^{189}\mathrm{Os}$, demonstrating measurable detection rates of the NEEC process and discussing the feasibility of the proposed approach. This new approach for observing the NEEC process may be realized in the near future.

\end{abstract}

\maketitle

A long-lived isomer is a distinct metastable quantum state of a nucleus ~\cite{Aprahamian2005}, with potential applications in nuclear energy storage ~\cite{Walker1999} and nuclear clocks ~\cite{Kraemer2023}. Over the past few decades, significant efforts have been directed toward investigating the feasibility of exciting the isomeric state assisted by electrons from the atomic shell. One promising mechanism is nuclear excitation by electron capture (NEEC), which is the reverse process of internal conversion (IC). In the NEEC process, an electron recombines into a vacancy of an ionic atom, and simultaneously the nucleus is excited to a higher energy state. This only occurs when the energy released from the electron capture matches the excitation energy of the nucleus.

After NEEC was initially predicted in 1976 ~\cite{GOLDANSKII1976}, numerous proposals were made for observations of NEEC in various setups, including plasmas ~\cite{Gunst2014,PhysRevC.59.2462,PhysRevLett.120.052504,Gunst2018,PhysRevLett.130.112501,Rom.Rep.Phys.68.S37,J.Phys.G45.033003}, electron beam ion traps (EBITs)~\cite{Wang2023,DillmannWorkshop2022,ZhaoarXiv2024}, crystals~\cite{Europhys.Lett,Morel2005}, and storage rings~\cite{Pálffy2008,ZhaoarXiv2024}. So far, the NEEC observation is very challenging, and the sole evidence of NEEC observation was reported through a beam-based experiment in the isomer depletion of $^{93m}\mathrm{Mo}$~\cite{Chiara2018}. However, the measured excitation probability significantly exceeds theoretical expectations~\cite{Wu2019,PhysRevLett.127.042501,PhysRevC.108.L031302}, and discussions on the background consideration have also arisen on the experimental side~\cite{Guo2021,Chiara2021}. Subsequently, a similar measurement under low $\gamma$-ray background conditions was performed. However, no signal of $^{93m}\mathrm{Mo}$ isomer depletion was observed~\cite{Guo2022}. The key factor impeding conclusive observations of NEEC lies in the competing nuclear excitation processes such as photoexcitation and inelastic scattering~\cite{Chiara2018,Gunst2018,PhysRevLett.130.112501}, as well as pervasive background noise~\cite{Guo2021,Chiara2021} such as x-/$\gamma$-rays from collisions and atomic processes. 

In this Letter, we introduce a novel experimental approach to confront these challenges, aiming to observe the NEEC process by combining an EBIT as site for the NEEC process and Penning-trap mass spectrometry (PTMS) for the detection step. The EBIT can produce a quasi mono-energetic electron beam with a tunable electron energy up to 200 \SI{}{\kilo\electronvolt} and an energy spread of a few tens \SI{}{\electronvolt}~\cite{Beiersdorfer1996, UTTER1999, Crespo2004, Kuramoto2002, Fu2010}. These are ideal experimental conditions with  well-defined initial and final atomic and nuclear states for the occurrence of  NEEC. Following excitation, the long-lived nuclear isomers could be extracted from the EBIT and transferred to the Penning trap apparatus for isolation and interrogation. The latter is a powerful tool in the determination of atomic masses with the highest precision, and enables the detection of nuclear isomeric states~\cite{Block2008,Gallant2012,Nesterenko2020} or even electronic metastable states~\cite{Schüssler2020,Kromer2023} through the measurement of cyclotron frequencies of the charged particles trapped in a strong magnetic field. Thus, through specific NEEC resonances to generate long-lived isomers in an EBIT, unambiguous mass measurements can be performed to identify the isomers via PTMS, eliminating the need for fluorescence photon detection or IC electron detection which would be strongly affected by pervasive background. 
The EBIT and PTMS combination promises measurable detection rates and signals which can be unambiguously assigned to the NEEC process, providing also experimental cross-section values. This method would usher in a new platform to observe NEEC, offering clear verification of the NEEC phenomenon and theory.

We focus on the interesting candidate of $^{189}\mathrm{Os}$. 
This nucleus hosts a high-spin, long-lived isomer $9/2^{-}$ (denoted here as $^{189m}\mathrm{Os}$) located \SI{30.82}{\kilo\electronvolt} above the nuclear ground state, commensurable with atomic state energies. The half-life is about $5.8$ hours for neutral atoms, while it is more than $206.6$ years for highly charged ions (HCIs), such as $^{189m}\mathrm{Os}^{74+}$, due to the lack of IC channels. Thus, it is a suitable nucleus to observe the NEEC process using the proposed method. By conducting a theoretical calculation of the NEEC cross section based on the method developed in Refs.~\cite{Palffy2006,Gunst2015}, and population dynamics of the long-lived isomer, our results strongly support the feasibility of detecting the NEEC process within a background-free environment.



\begin{figure}[t!]
\centering
\resizebox{0.45\textwidth}{!}{
\includegraphics{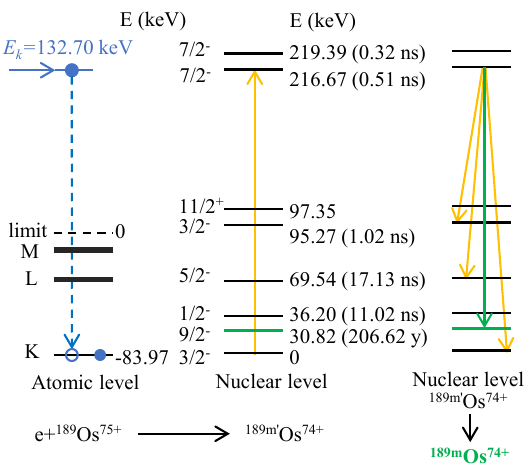}}
\caption{\justifying Atomic and nuclear level diagrams illustrating the nuclear excitation by electron capture process from $^{189}\mathrm{Os}^{75+}$ to $^{189m}\mathrm{Os}^{74+}$, which includes two steps. In the first step, a free electron with kinetic energy $E_k$ is captured by $^{189}\mathrm{Os}^{75+}$ and the nucleus is excited to a higher lying nuclear state $^{189m'}\mathrm{Os}^{74+}$. The second step is the decay of $^{189m'}\mathrm{Os}^{74+}$ to the lower states. The energy levels of electrons and nuclei are obtained from Ref.~\cite{NIST_ASD} and Ref.~\cite{ENSDF}, respectively. The lifetime of nuclear levels is determined by the spontaneous decay for He-like ions.}
\label{figOslev:res}
\end{figure}

\begin{figure*}[t]
\centering
\resizebox{0.9\textwidth}{!}{
\includegraphics{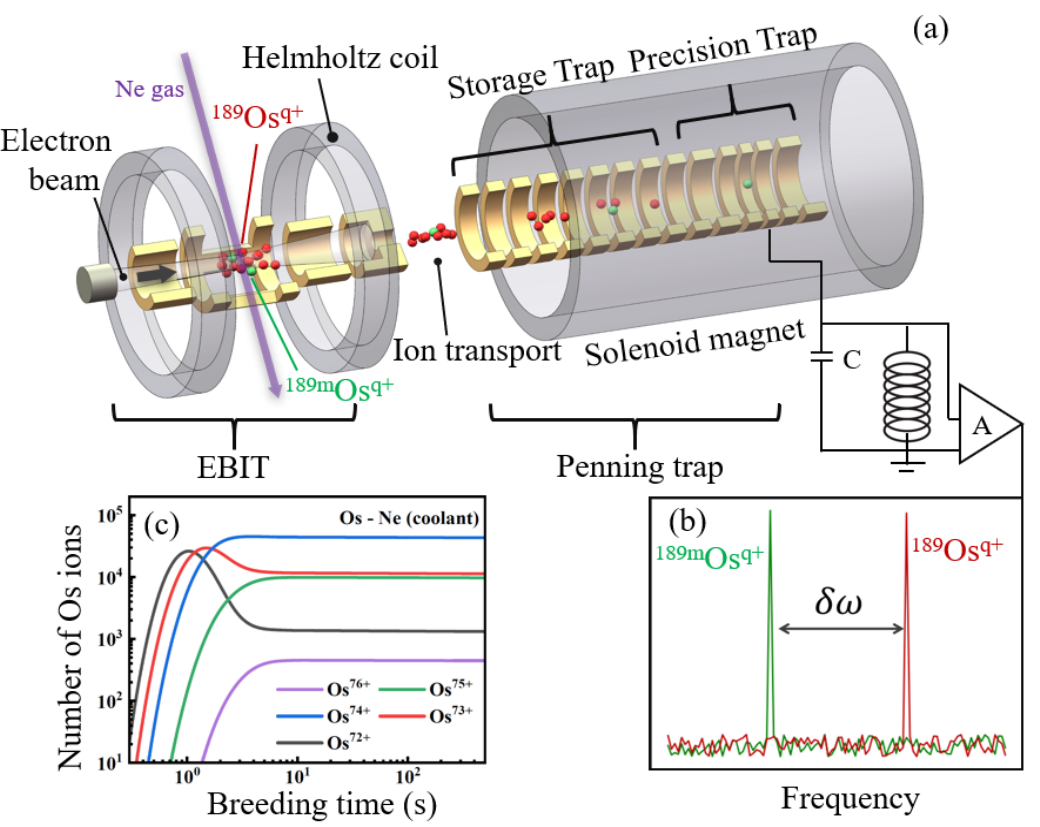}
} 
\caption{\justifying The schematic diagram of the proposed experimental setup consists of an EBIT and a Penning trap. (b) Simulated fast Fourier transform (FFT) spectrum of the signals from two ions in the nuclear ground (red) and isomeric state (green), respectively. 
(c) Simulated charge state evolution of Os ions with $q=72$ to $q=76$. The Os atoms are injected in a pulsed mode while Ne gas are continuously injected as a coolant. Further details regarding the EBIT settings can be found in the main text.}
\label{figSetup:res}
\end{figure*}

In the process of NEEC, a free electron recombines with an ion to form a bound state under release of energy, ultimately leading to the excitation of the nucleus. We follow the method developed in Refs.~\cite{Palffy2006,Gunst2015} to calculate NEEC cross sections. For a specific capture channel $\alpha$, the NEEC cross section as a function of the electron kinetic energy $E_k$ can be expressed as 
\begin{equation}
\sigma_{\mathrm{NEEC}}^{\alpha} (E_k)=\frac{2\pi^{2}  }{p^{2} } Y_{\mathrm{NEEC}}^{\alpha}L_{d} (E_k-E_{d} ) .
\end{equation}
Here, $p$ represents the momentum of the free electron and $Y_{\mathrm{NEEC}}^{\alpha}$ is the NEEC transition rate~\cite{Gunst2015}. The function $L_{d}$ represents the normalized Lorentz profile occurring in resonant systems,
\begin{equation}
L_{d} (E_k-E_{d} )=\frac{\Gamma _{d}/2\pi }{(E_k-E_{d} )^{2}+\Gamma _{d}^{2}/4}  ,
\end{equation}
with $E_{d}$ and $\Gamma _{d}$ being the energy and the natural width of the resonant state, respectively. The width $\Gamma _{d}$ of the NEEC resonance for a given NEEC channel is the sum of the nuclear state width and the electronic width. The resonance strength of NEEC is determined by integrating the reaction cross section over electron energy $S_{\mathrm{NEEC}}^{\alpha } =\int dE_k\sigma _{\mathrm{NEEC}}^{\alpha } (E_k)$.

The candidate scenario for probing NEEC process in the $^{189}\mathrm{Os}^{75+}$ ion is depicted in Fig.~\ref{figOslev:res}. The $^{189}\mathrm{Os}^{75+}$ gets excited to the nuclear level $7/2^{-}$ at \SI{216.67}{\kilo\electronvolt} by capturing a free electron with kinetic energy $E_k =  \SI{132.70}{\kilo\electronvolt}$. Subsequently, the excited nucleus can decay to its long-lived isomeric state $9/2^{-}$ at \SI{30.82}{\kilo\electronvolt} with a specific branching ratio $B_r$. This process bears resemblance to the pumping of nuclear isomeric states with synchrotron radiation or high-power lasers~\cite{Nature573.238,PhysRevLett.128.052501}. In principle, the long-lived isomeric state could be directly excited by capturing a free electron with lower kinetic energy to higher electronic orbitals. However, the cross section for such direct excitation to the long-lived isomeric state is very low according to our calculation. Furthermore, calculations following the method in Ref.~\cite{ZhangPRC2022} show that the direct excitation to the long-lived isomeric state from the ground state by inelastic electron scattering is also negligible, under the conditions analysed in the present work.

As the NEEC channel for the indirect excitation to the isomeric state studied in the present work involves only electronic ground states, the width $\Gamma_{d}$ of the NEEC resonance is the nuclear state width which is on the order of $10^{-6}$ eV for the concerned $7/2^{-}$ level. Due to the narrow width of the resonant state, the yield of isomeric states for a single nucleus through the capture orbital $\alpha$ is
\begin{equation}
\begin{split}
A_\mathrm{NEEC}&=B_r\int dE_k\sigma _{\mathrm{NEEC}}^{\alpha } (E_k)\phi _{e} (E_k,n_{e},\Gamma _{e}) \\
&\approx B_rS_{\mathrm{NEEC}}^{\alpha }\phi _{e} (E_d,n_{e},\Gamma _{e}),
\label{Aneec}
\end{split}
\end{equation}
where $n_{e}$ represents the electron number density, $\Gamma _{e}$ stands for the Gaussian distribution width, and $\phi _{e}$ is the electron flux distribution. In the present work, electron wavefunctions from GRASP2K~\cite{JONSSON20132197} and nuclear reduced transition probabilities from Ref.~\cite{ENSDF} are employed for the calculation of NEEC cross sections.


The envisaged experiment can be performed involving a combination of an EBIT and a Penning trap, as shown in Fig.~\ref{figSetup:res}. In the EBIT, the electron beam is produced by an electron gun and then accelerated to high energy using a bias voltage supply. A pair of Helmholtz-shaped coils generates a homogeneous magnetic field with a strength of a few tesla along the axis of the drift tube assembly, which can compress the electron beam, resulting in a high electron density. The Os atoms can be injected into the EBIT in pulsed mode from external sources. After continuous collisions with the electron beam, Os atoms become highly charged and axially trapped by the trap depth, and in radial direction they are confined by a combination of magnetic field and space-charge potential produced by the electron beam. By tuning the beam energy to the resonance of NEEC, the isomers could be generated in the EBIT. A transmission beam line is employed to select the charge state and transport the ions from the EBIT to the Penning trap apparatus, as already implemented in Refs.~\cite{Schüssler2020,Morgner2023}. Once the batch of ions is captured by the Penning trap, by employing multi-potential wells, the ions can then be separated incrementally until a single ion is isolated for cyclotron frequency measurement meanwhile the remaining ions are still stored -- a standard procedure demonstrated in Penning trap experiments~\cite{sturm_alphatrap_2019,Smorra_base_2015}. The isolated ions are then transported to a precision trap for cyclotron frequency measurements, one by one. With a typical magnetic field of \SI{7}{\tesla}, the cyclotron frequency of $^{189}\mathrm{Os}^{74+}$ is about 42 MHz, while the frequency of its long-lived isomer $^{189m}\mathrm{Os}^{74+}$ is approximately \SI{7}{\hertz} lower, since the equivalent atomic mass is heavier for the isomer according to the mass–energy equivalence. A standard measurement of the modified cyclotron frequency to distinguish the isomer can be completed within several tens of seconds using double dip spectra~\cite{sturm_alphatrap_2019}, or even faster with the phase-sensitive measurement~\cite{sturm_phase-sensitive_2011}. Since the atomic processes in EBIT such as electron collision ionization (CI), radiative recombination (RR) and charge exchange (CX) can not excite nuclear states, the detection of an isomer in the Penning trap provides unambiguous evidence of the occurrence of NEEC. 


Given the achievable EBIT specifics including a \SI{300}{\milli\ampere} electron beam current, a beam radius of \SI{40}{\micro\meter}, a beam energy of \SI{132.70}{\kilo\electronvolt} and $\Gamma_{e} = \SI{30}{\electronvolt}$, the isomer generation rate $A_\mathrm{NEEC}$ can be calculated to be \SI{1.35e-7}{\per\second}. This calculation uses the theoretical NEEC resonance strength of \SI{1.05e-3}{\barn\eV} for the $7/2^{-}$ states at \SI{216.67}{\kilo\electronvolt}. The branch ratio $B_r$ is 25.9$\%$, obtained by calculating the radiative decay rates and internal conversion de-excitation rates for each decay channel. The low rate requires a long isomer breeding time in the EBIT and a fine control over ion loss. Given that the trapped ions are continually heated by electron collisions, leading to thermal loss, ion cooling becomes imperative. One of the most effective methods is the so-called sympathetic evaporative cooling. This is achieved by continuously injecting a lighter cooling gas, such as neon, which takes energy from the highly charged ions, allowing them to remain trapped while the light ions, serving as coolants, preferentially escape~\cite{Penetrante1991-2}. This approach was successfully demonstrated by Schneider \textit{et al}~\cite{Schneider1989} for cooling $\mathrm{Au}^{69+}$ ions. They observed an exponential decay time of 4.5 hours for the $\mathrm{Au}^{69+}$ signal, despite the presence of heavier lead ion accumulation in the trap, which can diminish the cooling effect on gold. In the case of a higher charge state of Os and in the preparation of a lead-free environment, a lower ion escape rate can be anticipated. 

In this study, we employed the theoretical model established by Penetrante \textit{et al}~\cite{Penetrante1991,Penetrante1991-2} to calculate the charge state abundance and the rate of ion loss for highly charged Os ions being evaporatively cooled by a Ne gas. In our calculations, we considered atomic processes such as CI, RR and CX, while neglecting at this stage  nuclear processes due to their significantly lower reaction rates. The heating rate of ions resulting from collisions with the electron beam as well as the energy transfer between Os and Ne ions was calculated to determine the ion escape rates. By solving the rate equations, we calculated the evolution of charge state abundance, revealing that equilibrium is typically reached within approximately \SI{10}{\second}, see Fig.~\ref{figSetup:res}(c). Our calculations also underscored a strong dependence on the density of the cooling gas. At steady state, with $6 \times 10^4$ highly charged Os ions and a Ne gas density of \SI{2.5e5}{\centi\meter^{-3}}, the average ion escape rate for the relevant charge state is $1.5 \times 10^{-5}$ \SI{}{\second^{-1}}. Further details are presented in Supplemental Material~\cite{Supp.mat}. 


\begin{figure}[th]
\centering
\resizebox{0.45\textwidth}{!}{
\includegraphics{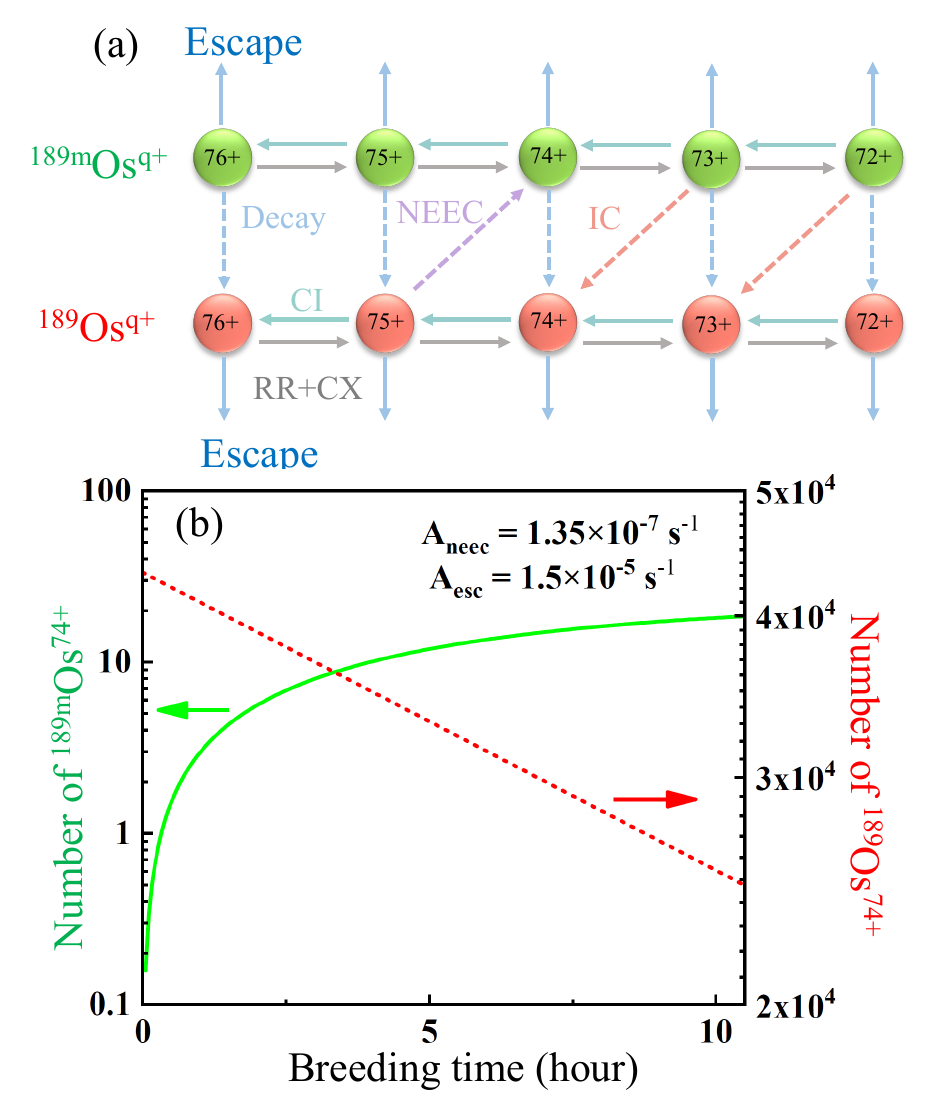}
}
\caption{\justifying (a) The plasma model established for the isomer breeding calculation involving the atomic processes such as CI, RR, CX and nuclear processes of NEEC, IC and spontaneous decay (labeled ``Decay"), as well as the ion escape for $^{189}\mathrm{Os}^{q+}$ and $^{189m}\mathrm{Os}^{q+}$ with the dominant charge states of $q=72$ to $q=76$. (b) Simulated total number of Os ions (dash line) as well as the isomers (solid line) as a function of breeding time in the EBIT. The averaged ion escape rate is \SI{1.5e-5}{\second^{-1}} and isomer generation rate through the NEEC is \SI{1.35e-7}{\second^{-1}}.} \label{fig:number}
\end{figure}

To assess the quantity of the long-lived isomers that can be generated, we establish a plasma model for isomer breeding involving both atomic and nuclear processes, as depicted in Fig.~\ref{fig:number}(a), which can be solved by a set of multi-charge-state rate equations,
\begin{equation}
\begin{aligned}
\frac{dN_{q,\, g}}{dt} = & N_{q+1, \,g\,} (A_{\mathrm{RR}{,\, q+1 \rightarrow q}} + A_{\mathrm{CX}{,\, q+1 \rightarrow q}}) \,  \\ 
& +  N_{q-1,\, g\,} A_{\mathrm{CI}{,\, q-1 \rightarrow q}} + N_{q,\, m} A_{\mathrm{Dec},\,m \rightarrow g,\, q} \, \\
&  + N_{q-1,\, m}A_{\mathrm{IC},\,m \rightarrow g,\, q-1} - N_{q,\, g\,} (A_{\mathrm{CI}{,\, q \rightarrow q+1}} \\
& + A_{\mathrm{RR}{,\, q \rightarrow q-1}} \, +  A_{\mathrm{CX}{,\, q \rightarrow q-1}}  + A_{\mathrm{esc}{, \, q}} \\
& + A_\mathrm{NEEC,\, q }) \\
\frac{dN_{q,\, m}}{dt} = & N_{q+1, \,m\,} (A_{\mathrm{RR}{,\, q+1 \rightarrow q}} + A_{\mathrm{CX}{,\, q+1 \rightarrow q}}) \,  \\
& + N_{q-1,\, m\,} A_{\mathrm{CI}{,\, q-1 \rightarrow q}} + N_{q+1,\, g \,} A_\mathrm{NEEC,\, q + 1 } \,  \\
& - N_{q,\, m\,} (A_{\mathrm{CI}{,\, q \rightarrow q+1}} + A_{\mathrm{RR}{,\, q \rightarrow q-1}} \, \\
&+ A_{\mathrm{CX}{,\, q \rightarrow q-1}}  + A_{\mathrm{Dec},\,m \rightarrow g,\, q}+A_{\mathrm{IC},\,m \rightarrow g,\, q}\\
&+ A_{\mathrm{esc}{, \, q}}), 
\label{eq:x}
\end{aligned}
\end{equation}
\noindent where $A_{\mathrm{RR}}$, $A_{\mathrm{CX}}$ and $A_{\mathrm{CI}}$ represent the RR, CX and CI rates, respectively, calculated according to Ref.~\cite{Penetrante1991}. $A_{\mathrm{Dec}}$ and $A_{\mathrm{IC}}$ denote the spontaneous decay and internal conversion rates, respectively. $A_{\mathrm{esc}{,\,q}}$ is the escape rate of ions for charge $q$. The isomer can be produced through NEEC from $^{189}\mathrm{Os}^{75+}$ to $^{189m}\mathrm{Os}^{74+}$. After $^{189m}\mathrm{Os}^{74+}$ is produced via NEEC, it will appear in the relevant charge state immediately via the atomic processes such as CI and RR. The IC rates are about \SI{2.45e-6}{\per\second} and \SI{4.9e-6}{\per\second} only for $q = 73$ and $q=72$ due to the opening of IC channels. The spontaneous decay of the isomer can be neglected due to the highly forbidden transition.  

In Fig.~\ref{fig:number}(b), the simulation result demonstrates that approximately 18 isomers, primarily $^{189m}\mathrm{Os}^{74+}$, can be generated after a breeding duration of 10 hours in the EBIT. By transporting all ions ($N\approx 2.5\times 10^4$) with a selected charge of $q=74$ downstream to the Penning trap, we can effectively store them and perform individual mass measurement to accurately quantify of ions in both their ground and isomeric states. This allows for the evaluation of the isomer population with the specified breeding duration, expressed as 
%
\begin{equation}
\mathrm{P}(t)\approx 1-\exp\left({-\frac{N_\mathrm{75+}}{N_\mathrm{total}}A_\mathrm{NEEC}\,t}\right),\label{Pisomer}
\end{equation}
where $N_\mathrm{75+}/N_\mathrm{total}$ represents the fractional ion abundance for $q=75$. Note that this approximation is applicable when the breeding duration does not exceed the inverse of the ion escape rate and when the number of the isomers remains significantly lower than the total ion number. In terms of Eq.~\eqref{Aneec}, the resonance strength $S_{\mathrm{NEEC,expt.}}^{\alpha }$ can be determined with a measured $\mathrm{P}(t)$, 
\begin{equation}
S_{\mathrm{NEEC,expt.}}^{\alpha }=-\frac{N_\mathrm{total}}{N_\mathrm{75+}}\frac{\mathrm{ln}(1-\mathrm{P}(t))}{B_r \phi _{e} (E_d,n_{e},\Gamma _{e}) t },
\end{equation}
where the electron density can be measured by a slit imaging experiment~\cite{UTTER1999}. The charge-state distributions as well as the number of ions can be obtained via ion extraction~\cite{Penetrante1991}. It is worth noting that the electron beam energy is determined by the nominal beam energy, defined by the voltage applied to the EBIT, along with a modification arising from the space-charge effect. The space-charge effect can be characterized through dielectronic recombination (DR) measurements, a method widely adopted by many EBIT groups~\cite{Yang2023,Martinez2006}. Alternatively, radiative recombination measurements can also be used to analyze this effect~\cite{Marrs1995}. Meanwhile, the energy spread can be determined using resonant recombination measurements~\cite{Beilmann2009}. 

The total time spent on the entire measurement campaign depends on both the time required for mass measurement of each Os ion and the probability of isomers production in the EBIT. According to our calculations, the anticipated probability, based on the predicted NEEC cross section from theory, is 0.07\% over a 10-hour period. During the transport to the Penning trap, those ions can be lost, however, this doesn't alter the isomer probability to be measured. Assuming an achievable transmission efficiency of 2\%, 500 ions can be captured and stored with a sufficiently long time in storage traps. With a reasonable measurement time for a single $^{189,189m}\mathrm{Os}^{74+}$ ion, 500 running cycles can be completed in 10 hours, and consequently the successful detection of a single isomer, even with a 0.07\% isomer probability, can be achieved within two days. Given that the resonance energy is known with an uncertainty of approximately $20$ eV \cite{NIST_ASD,ENSDF}, a complete resonance scan can be conducted using a 30-eV wide electron beam. Based on our Monte Carlo simulations, this approach is expected to yield approximately 28 detected NEEC events over 60 days, corresponding to a statistical uncertainty of around 20\%. To improve the statistics, it is essential not only to accelerate the ion mass measurement but also to effectively control ion escape in the EBIT. With an even lower loss rate, an extended trapping time can be employed, increasing the probability of detecting the isomer. 


In conclusion, we propose a novel idea to detect the NEEC process via a combination of an EBIT and an Penning trap. In the EBIT, the long-lived isomeric state $9/2^{-}$ at \SI{30.82}{\kilo\electronvolt} can be excited through NEEC that occurs by collisions with an electron beam target at resonant energy. Subsequently, the isomer can be identified by precision mass measurements using the PTMS. In  comparison with other proposals via spectroscopic measurements, the advantage of our approach lies in the unambiguous detection of the isomer, which serves as definitive evidence for the occurrence of NEEC. We believe that this experimental approach, based on precision mass spectrometry, holds broad applicability, not only facilitating the observation of NEEC in other nuclei but also allowing for the investigation of other nuclear processes such as nuclear excitation by electronic bridge~\cite{Bilous2020}, nuclear excitation by electron transition~\cite{Morita1973,Chodash2016} and nuclear excitation by two-photon electron transition~\cite{Volotka2016}. These capabilities are drawing interest in various research fields in nuclear physics, including isotope separation, energy storage, and nuclear clocks. 
\begin{acknowledgments}
We would like to thank Ziwen Li for helpful discussions. This work is supported by the National Key Research and Development Program of China under Contract No. 2024YFA1610900, No. 2022YFA1602504 and No. 2023YFA1606501, and the National Natural Science Foundation of China (NSFC) under Contract No. 12447106, 12147101, No. 12475122, No. 12204110, and No. 11874008, Shanghai Pujiang Program under Grant No. 22PJ1401100 and Max-Planck Partner Group Project. A.P. gratefully acknowledges support from the Deutsche Forschungsgemeinschaft (DFG, German Science Foundation) in the framework of the Heisenberg Program (PA 2508/3-1).
\end{acknowledgments}

\bibliographystyle{unsrtnat}
\bibliography{reference}  


\end{document}